\documentclass{dcds2}

\def\ppages{1--10}

\usepackage{amssymb}
\theoremstyle{plain}

\theoremstyle{definition}

\newcommand{\dif}{\mathrm{d}}
\newcommand{\expe}{\mathrm{e}}
\newcommand{\Hamil}{{\mathcal H}}

\newcommand{\Espec}{E}
\newcommand{\nspec}{n}
\newcommand{\Escale}{E^{(\varepsilon)}}
\newcommand{\Eren}{\bar{E}}
\newcommand{\gwid}{\beta}
\newcommand{\gtime}{\gamma}
\newcommand{\Intexpt}{{\mathcal K^{(t)}}}
\newcommand{\Intexptren}{{\mathcal K^{(\tren)}}}
\newcommand{\sgn}{\,\mathrm{sgn}\,}
\newcommand{\kdisc}{k}
\newcommand{\kdisca}{\kdisc_1}
\newcommand{\kdiscb}{\kdisc_2}
\newcommand{\kdiscc}{\kdisc_3}
\newcommand{\kcont}{\xi}
\newcommand{\kconta}{\xi_1}
\newcommand{\kcontb}{\xi_2}
\newcommand{\kcontp}{\xi^{\prime}}
\newcommand{\kren}{\bar{\xi}}
\newcommand{\krena}{\kren_1}
\newcommand{\krenb}{\kren_2
}
\newcommand{\tren}{\bar{t}}
\newcommand{\omegaa}{\omega_{\kdisca}}
\newcommand{\omegab}{\omega_{\kdiscb}}
\newcommand{\omegac}{\omega_{\kdiscc}}

\newcommand{\amode}{a}
\newcommand{\amodekd}{a_{\kdisc}}
\newcommand{\saa}{s_1}
\newcommand{\sbb}{s_2}
\newcommand{\scc}{s_3}
\newcommand{\Hcoup}{\Hamil}
\newcommand{\Ifin}{L}
\newcommand{\Ecoup}{\tilde{L}}
\newcommand{\BBR}{\mathbb{R}}
\newcommand{\BBZ}{\mathbb{Z}}
\newcommand{\Circ}{S_1}

\newcommand{\intd}[2]{\int_{#1}^{\sharp(#2)}}
\newcommand{\deldisc}[1]{\tilde{\delta}^{\sharp(#1)}}
\newcommand{\ord}{\,\mathrm{ord}\,}
\newcommand{\tlin}{t_{\mathrm{L}}}
\newcommand{\tnl}{t_{\mathrm{NL}}}
\title[Application of Weak Turbulence Theory to FPU Model]
 {Application of Weak Turbulence Theory to FPU Model}

 \email{kramep@rpi.edu}

\author[Peter R. Kramer, Joseph A. Biello, and Yuri V. Lvov]{}

\begin{document}
\maketitle

\centerline{\scshape Peter R. Kramer, Joseph A. Biello, and Yuri Lvov }
\medskip

{\footnotesize \centerline{} 
\centerline{Department of Mathematical Sciences}
\centerline{Rensselaer Polytechnic Institute}
 \centerline{301 Amos Eaton Hall}
 \centerline{110 8th Street}
 \centerline{Troy, NY 12180, USA}
 }

\begin{quote}{\normalfont\fontsize{8}{10}\selectfont
{\bfseries Abstract.}
The foundations of weak turbulence theory is explored through its
application to the (alpha) Fermi-Pasta-Ulam (FPU) model, a 
simple weakly nonlinear dispersive system.  A direct application of
the standard kinetic equations would miss interesting
dynamics of the energy transfer process starting from a large-scale
excitation.  This failure is traced to an enforcement of the exact
resonance condition, whereas mathematically the resonance should be
broadened due to the energy transfer happening on large but finite
time scales.  By allowing for the broadened resonance, a modified
three-wave kinetic equation is derived for the FPU model.  This
kinetic equation produces some correct scaling predictions about the
statistical dynamics of the FPU model, but does not model accurately
the detailed evolution of the energy spectrum.  The reason for the failure 
seems not to be one of the previously clarified reasons for breakdown
in the weak turbulence theory.
\par}
\end{quote}

\section{Introduction}
Weak turbulence (WT) theory was developed in the 1960s to provide
equations which quantitatively describe the transfer of energy among
turbulent, 
weakly nonlinear, dispersive waves in
fluids~\cite{djb:nlrwd,jb:rwc,kh:onlet1,bbk:pt,acn:wti,acn:wtaai,vez:kst1}.
The basic ``kinetic equations'' produced by WT theory have
been applied to analyze  energy transfer
include internal and surface waves with small
aspect ratios in the atmosphere and ocean~\cite{yvl:hfgms,vez:kst1},
semiconductor lasers~\cite{yvl:qwtas}, and plasma
turbulence~\cite{bbk:pt}.
The domain of validity of
the governing equations, however, remains to be fully explored.  

Zakharov~\cite{vez:kst1} and Kadomtsev~\cite{bbk:pt} obtained the
kinetic equations through arguments (such as random phase approximation)
motivated by physical intuition, whereas various authors in the USA and Western
Europe~\cite{djb:nlrwd,jb:rwc,kh:onlet1,acn:wti} 
derived the same kinetic equaitons through a multiple scales
expansion.  While the latter approach is systematic, it has not been
rigorously formulated, and several breakdowns of its validity have
been noted in recent years.  Newell and coworkers~\cite{ljb:sfcbc,cc:dawt,acn:wtaai} have noted
that the basic asymptotic expansion parameter is usually not uniformly
small over the full range of scales in a weakly turbulent system, so
that ``intermittent'' effects not directly predictable by WT theory
will manifest themselves at either sufficiently large or small
scales.  Cai, Majda, McLaughlin, and Tabak~\cite{dc:sbdwt,dc:dwtod,ajm:odmdw} 
have scrutinized the
validity of the kinetic equations of WT theory and their underlying
assumptions through their application on a simple one-dimensional
model amenable to comparison with direct numerical simulation.  One
key finding of their work is that under certain circumstances,
a nonlocal coupling between the
large-scale forcing and much smaller ``inertial range'' scales can
arise which invalidates the tacit assumption that the energy transfer
in this inertial range of scales is unaffected by the presence of
forcing and/or damping.  

To further these efforts at understanding the domain of validity of WT
theory, we study its application to a simple Hamiltonian model
without driving or damping. 
The Fermi-Pasta-Ulam (FPU) model of a
one-dimensional 
collection of particles coupled by weakly anharmonic springs serves as
a natural test example which ostensibly should be amenable to the WT
approach, provided the number $ N$ of particles in the chain is large
enough.  Indeed, the FPU model has dispersive waves which are coupled
through a weak nonlinearity, and it can be initialized with a
disordered wave profile.  Moreover, its Hamiltonian nature excludes
the problem related to the nonlocal force coupling.  
We consider here the $ \alpha $ version of the FPU model (defined in
Section~\ref{sec:fpumod}) 
which has a cubic nonlinearity in the Hamiltonian and therefore direct
three-wave coupling (which is usually simpler to analyze than 
four-wave couplings as studied in~\cite{dc:sbdwt,dc:dwtod,ajm:odmdw}).
Nonetheless, we still find the standard kinetic equations
of weak turbulence insufficient to predict the dynamics of the energy
spectrum in the FPU model when it is initialized with a large-scale
excitation.  

As we discuss in Section~\ref{sec:modkin} after some notational
preparation in Section~\ref{sec:not},  
there are no nontrivial exact triad (three-wave)
resonances.  A literal
application of the standard WT theory would suggest then that the energy
transfer between waves must be governed by four-wave interactions.  A
scrutiny of the numerical simulations of the $ \alpha $-FPU
dynamics~\cite{jab:setfp},
however, indicates the presence of at least two distinct time scales on which
significant energy transfer occurs.  The first phase only involves
energy transfer and equilibration 
among modes of sufficiently small wavenumber.  In the
final phase, occurring on a much longer time scale, 
energy is redistributed among all modes.  These observations can be
reconciled with the resonance properties of the dispersion relations
by recognizing that the low wavenumber modes are in \emph{near
resonance} because the waves become approximately acoustic at low
wavenumber.  Consequently, we can interpret the first phase of energy
transfer among low wavenumber modes as being mediated by triad interactions and
the later phase of relaxation to global energy equipartition as
requiring quartet interactions. 

We attempt here to adapt weak turbulence theory to describe
the energy transfer among nearly resonant triads during the earlier
phase of evolution of the energy spectrum in the $ \alpha$-FPU model.
Revisiting the multiple scales derivation of the three-wave kinetic
equation~\cite{acn:wti} in Section~\ref{sec:derkin}, 
we identify a step in which a large but finite time
is replaced by an infinite time limit. It is this simplification which
enforces the exact resonance condition present in the standard kinetic
equations.  In reality, the resonance should be broadened simply
because the three-wave kinetic equation seeks to describe the
statistical dynamics of a turbulent system on at a large but finite
time.  Over a long but finite time $  T $, frequency mismatches of
order $ o(T^{-1}) $ cannot be detected.  By properly retaining the
effects of this ``finite-time resonance broadening'' and employing a
systematic renormalization group technique, we can compute the range
of wavenumbers which will exchange significant energy through triad
interactions and the time scale on which this energy exchange should
occur.  These scaling predictions are well confirmed by the numerical
results, and agree with earlier partial
predictions~\cite{bc:stfpu,lg:stsco,dls:lecfp}.

Our calculations also produce a modified kinetic equation
(\ref{eq:modkin}) with
finite-time resonance broadening.  
Unfortunately, the details of this
kinetic equation are not in good agreement with the numerically
observed energy spectrum evolution.  Moreover, the modified kinetic
equation generates negative energies.  The problem seems to be the
omission of higher order secular terms in the multiple scales
expansion which play an important role when the waves are weakly
dispersive~\cite{vsl:sdat}, but does not appear to fall neatly into an
established situation where the WT theory is known to fail.
In future work, we will attempt to clarify the nature of the breakdown
and assess whether a further modified WT approach might succeed.
\section{Fermi-Pasta-Ulam Model}
\label{sec:fpumod}
The Fermi-Pasta-Ulam (FPU) model is a model for a one-dimensional 
collection of particles with
massless, weakly anharmonic (nonlinear) springs connecting them to each other.
Letting $ \{q_j\}_{j=1}^{N} $ and $ \{p_j\}_{j=1}^{N} $ denote the
position and momentum coordinates of an $N$-particle chain, we can
define the ($\alpha $ version) FPU model Hamiltonian, which we write
immediately in nondimensional form:
\begin{eqnarray}
\Hamil&=&\frac{1}{2}\sum\limits_{j=1}^N \left( {p_j^2} +
(q_j-q_{j+1})^2\right)+\frac{\epsilon}{3}
\sum\limits_{j=1}^N
(q_j-q_{j+1})^3\label{HamPhys}\\
\frac{\dif q_j}{\dif t} &=& p_j \nonumber\\
\frac{\dif p_j}{\dif t} &=& 
\left(q_{j-1}-2 q_j + q_{j+1} \right)\left(1+\epsilon(q_{j-1}-q_{j+1})\right)
\nonumber
\end{eqnarray}
The coordinates are normalized so that the total Hamiltonian energy
$\Hamil = N$, implying a typical $ O(1) $ value of each $
q_j $ and $ p_j $.  The small parameter $ \varepsilon \ll 1 $ then
measures the weakness of the nonlinearity.

We now make a conformal transformation to action-angle variables to
facilitate the application of WT theory.  We describe the
transformation by the following two steps:
\begin{eqnarray}
\left(\begin{array}{c}Q_k \\ P_k \end{array}\right)
&=&
\sum\limits_{j=1}^{N}
\left(\begin{array}{c}q_j \\ p_j \end{array}\right)
\exp{\left(\frac{2\pi i k j}{N}\right)}  \nonumber \\
\amodekd &=& \frac{1}{\sqrt{2 \omega_\kdisc}} \left(P_\kdisc - i \omega_\kdisc
Q_\kdisc\right),
\label{eq:discfour}
\end{eqnarray}
where $
\omega_\kdisc \equiv 2 \left|\sin{\left(\frac{\pi \kdisc}{N}\right)}\right| $
is the dispersion relation describing the frequency of a discrete
Fourier mode $ \kdisc $.  The virtue of the new (complex-valued) 
canonical variables $ \amodekd $ is that, in the absence of
nonlinearity, they rotate with a single frequency ( $ \amodekd \propto
\expe^{-i \omega_\kdisc t} $).  To connect with the standard notation of
WT theory and keep subsequent expressions as compact as possible, we
define the variables $
A_{\kdisc}^{-} \equiv \amodekd$, and $ A_{\kdisc}^+  \equiv
 \amode_{-\kdisc}^{*}$.
The superscript refers to the branch of the dispersion relation: $
A_{\kdisc}^{s} \propto \expe^{i s \omega_\kdisc t} $.

In these new variables, the Hamiltonian reads
\begin{eqnarray}
\Hamil&=&\frac{1}{N}
\sum\limits_{\kdisca,\kdiscb=-\frac{N}{2}+1}^{\frac{N}{2}}
\sum_{\saa,\sbb=\pm}
\Hcoup_{\kdisca,\kdiscb}^{\saa,\sbb} A_{\kdisca}^{\saa} A_{\kdiscb}^{\sbb}
\delta_{\kdisca+\kdiscb,0} \\
& & \qquad 
+ \frac{\varepsilon}{N^{2}}
\sum\limits_{\kdisca,\kdiscb,\kdiscc=-\frac{N}{2}+1}^{\frac{N}{2}}
\sum_{\saa,\sbb,\scc=\pm}
\Hcoup_{\kdisca,\kdiscb,\kdiscc}^{\saa,\sbb,\scc} A_{\kdisca}^{\saa}
A_{\kdiscb}^{\sbb} A_{\kdiscc}^{\scc}
\delta_{\kdisca+\kdiscb+\kdiscc,0}, \nonumber \\
\Hcoup_{\kdisca,\kdiscb}^{\saa,\sbb} &=& \frac{1}{2} 
\sqrt{\omega_{\kdisca} \omega_{\kdiscb}}
\delta_{\saa,-\sbb}, \\
\Hcoup_{\kdisca,\kdiscb,\kdiscc}^{\saa,\sbb,\scc}
&=& \frac{\saa \sbb \scc \sgn (\kdisca \kdiscb \kdiscc) 
\sqrt{\omegaa \omegab \omegac}}{6 \sqrt{2}}.
\label{eq:MostGeneral}
\end{eqnarray}
The symbol $ \delta_{i,j} $ denotes the Kronecker delta function.
% MEMO:  Make periodic
The equations of motion can then be written~\cite{acn:wti}
\begin{eqnarray}
\frac{\dif A_{\kdisc}^{s}}{\dif t} - i s \omega_{\kdisc} A_{\kdisc}^{s}
&=& \varepsilon \frac{1}{N} \sum_{\saa,\sbb=\pm} 
\sum\limits_{\kdisca,\kdiscb=-\frac{N}{2}+1}^{\frac{N}{2}}
\Ifin_{\kdisc \kdisca \kdiscb}^{s \saa \sbb} A_{\kdisca}^{\saa}
A_{\kdiscb}^{\sbb} \delta_{\kdisc,\kdisca+\kdiscb}, \\
\Ifin_{\kdisc \kdisca \kdiscb}^{s \saa \sbb} &=&
\frac{i \saa \sbb \sgn (\kdisc \kdisca \kdiscb) \sqrt{\omega_{\kdisc} \omegaa
\omegab}}{2 \sqrt{2}}. \label{eq:ifin}
\end{eqnarray}

To keep focus, we will restrict the initial data to have statistical
symmetry under parity inversion and to have Gaussian statistics.  
Then it suffices to describe
the energy transfer among different scales through the 
statistical quantities~\cite{acn:wti}: 
\begin{eqnarray}
\nspec_{\kdisc} \equiv \frac{1}{N} \langle |\amodekd|^2 \rangle
&=& \frac{1}{N} \langle A_{\kdisc}^{+} A_{-\kdisc}^{-} \rangle, \\
\Espec_{\kdisc} \equiv \omega_{\kdisc} \nspec_{\kdisc} + \omega_{-\kdisc} \nspec (- \kdisc) &=& 2 \omega_{\kdisc} \nspec_{\kdisc}. \nonumber
\end{eqnarray}
The main object of our interest will be $ \Espec_{\kdisc} $, 
the harmonic energy associated to the wavenumber $ \pm \kdisc $, or in
other words the mean contribution to the quadratic part of the Hamiltonian
in (\ref{eq:MostGeneral}) 
from wavenumbers $ \pm \kdisc $.  The mean-square amplitude
$ \nspec_{\kdisc} $ of the action variables is sometimes referred to
as the ``occupation number'' of the mode $ \kdisc $.  
\section{Notational Convention}
\label{sec:not}
The weak turbulence theory which we will apply to analyze the FPU
model is usually formulated in terms of spatially homogenous systems
of infinite extent, so that the Fourier wavenumbers are
continuously distributed.  This would be appropriate for a particular
well-defined  $ N
\rightarrow \infty $ limit of the FPU model, in which the physical lattice
spacing remains order unity and the system becomes infinitely extended
in both directions
in physical space.  We will always consider $ N$ to be finite but
large, and introduce notation which makes contact with that usually
used in the weak turbulence literature~\cite{acn:wti}.

We define the normalized wavenumber $ \kcont = \kdisc/N $ 
which takes values in the unit
circle $ \Circ $ (or equivalently, the space of real numbers $ \BBR $
with all numbers differing by an integer declared
equivalent)~\cite{DSWTrnb:ftia}.
To distinguish between Fourier coefficients defined with respect to
the integer wavenumbers $ \kdisc $ and the new wavenumber variable $
\kcont $, we write $ \amode_\kdisc $ for the former and $ \amode (\kcont) $ for
the latter. 
The associated frequency is defined through the
continuous dispersion relation $
\omega (\kcont) \equiv 2 \left|\sin \pi \kcont \right|$.

For large but finite $ N $, the
correlation functions of the Fourier coefficients are proportional to
a discretized delta-function, in the sense that
\begin{eqnarray*}
\langle \amode (\kcont) \amode^* (\kcontp) \rangle &=& \nspec (\kcont) 
\deldisc{N} (\kcont-\kcontp) \\
\deldisc{N} (\kcont) &\equiv& N \delta_{N\kcont,0}.
\end{eqnarray*}
where $ \nspec (\kcont) $ has a finite, nontrivial $ N \rightarrow \infty $
limit.  

To further emphasize the connection between the finite but large $ N $
and infinitely long systems, we introduce the discretized integral
notation:
\begin{displaymath}
\intd{D}{M} f (\kcont) \, \dif \kcont \equiv
\frac{1}{M} \sum_{\kcont: \kcont \in D \cap \BBZ/M} f(\kcont),
\end{displaymath}
which should just be thought of as a Riemann integral approximation to
the integral defined with respect to a mesh situated at integer
multiples of $ 1/M $.  

\section{Modified Kinetic Equation}
\label{sec:modkin}
The standard weak turbulence theory~\cite{acn:wti,vez:kst1} would predict that
three-wave interactions would be incapable of significant energy
transfer.  The reason is that the only triads of waves which satisfy
the exact resonance conditions
\begin{equation}
\kcont = \kconta + \kcontb, \qquad \qquad 
\omega (\kcont) = \pm \omega (\kconta) \pm
\omega (\kcontb)  \nonumber
\end{equation}
are those 
involving the zero mode ($ \kcont = 0 $ or $ \kconta = 0 $ or $
\kcontb = 0 $), which is uncoupled to the other degrees of
freedom due to the translational symmetry of the system
($\Ifin_{\kcont \kconta \kcontb} = 0$ if $ \kcont \kconta \kcontb =
0$).  The absence of three-wave resonances is here not due to the
finite size of the system~\cite{cc:dqwtc} but simply due to the
concavity of the dispersion relation~\cite{br:srpfp}.  

The reason the standard kinetic equations of weak turbulence theory
fail to describe any energy transfer through three-wave
interactions is that these interactions are effective only over a thin
band of wavenumbers $ \kcont $ with width vanishing with the
nonlinearity parameter $ \varepsilon $.  The derivation of the
standard weak turbulence kinetic equations assumes that all
quantities, including the wavenumbers $ \kcont $ of interest and the energy
density $ \Espec (\kcont) $ in each mode, are order unity in magnitude
(asymptotically independent of $ \varepsilon $).
The fact that the wavenumbers of interest during the three-wave
dynamics are small, and the energy density residing in these
wavenumbers large, requires a reworking of the asymptotic derivation of the
kinetic equation.  

To this end, a simple renormalization group approach is
useful~\cite{DSWTng:lptrg,ajm:smtd}.  We rescale the wavenumber and amplitude of
the energy spectrum by some gauge functions $ \gwid (\varepsilon) $
and $ \gtime (\varepsilon) $ depending on $ \varepsilon $:
\begin{equation}
\kren = \kcont/\gwid(\varepsilon), \qquad \qquad   
\tren = t/\gtime (\varepsilon). \label{eq:ktscale}
\end{equation}
The rescaling factors $ \gwid $ and $ \gtime $ are chosen so that $
\kren $ and $ \tren $ are order unity for the three-wave dynamics of
interest.   We shall leave them unspecified for now, 
other than insisting that $ \gwid \downarrow 0 $ and $ \gtime
\rightarrow \infty $ as $ \varepsilon \rightarrow 0 $, and show how
their scaling with $ \varepsilon $ 
can be determined 
by self-consistency.  More specifically, we define a
rescaled energy spectrum $ \Escale (\kren,\tren) $, obtained by
by applying the transformation (\ref{eq:ktscale}) to  the original energy spectrum:
\begin{equation}
\Espec (\kcont,t) = \gwid (\varepsilon)^{-1} 
\Escale (\kcont/\gwid (\varepsilon), t/\gtime (\varepsilon)) 
\label{eq:ektren}
\end{equation}
The rescaling of the amplitude results from the normalization
condition that the total quadratic energy is $ N$, 
which implies that if $ \Espec (\kcont,t) $ is
concentrated over a thin band of wavenumbers $ |\kcont| \lesssim
\gwid(\varepsilon) $, then the average amplitude of the energy spectrum 
over these modes must be $ \ord (\gwid(\varepsilon)^{-1}) $.  The
rescaled energy spectrum $ \Escale (\kren,\tren) $ is supposed to be
a truly order unity function, with order unity amplitude and 
order unity variations with respect
to $ \kren $ and $ \tren $ during the three-wave dynamical phase for
all sufficiently small $ \varepsilon$.  Mathematically, we demand that
the \emph{renormalized energy spectrum}
\begin{displaymath}
\Eren (\kren,\tren) = \lim_{\varepsilon \rightarrow 0} 
\Escale (\kren,\tren)
\end{displaymath}
be a nontrivial function with order unity amplitude and derivatives.
In renormalization group terminology, we want $ \Eren (\kren,\tren) $
to be a nontrivial fixed point under the transformation induced by
(\ref{eq:ektren}).  The
gauge functions $ \gwid (\varepsilon) $ and $ \gtime (\varepsilon) $
are determined by this property.  Indeed, as we shall show in
Section~\ref{sec:derkin}, 
\begin{equation}
\gwid = \varepsilon^{1/2}, \qquad 
\gtime = \varepsilon^{-3/2}. \label{eq:scalpred}
\end{equation}
and the renormalized energy spectrum satisifes the following modified
kinetic equation:
\begin{eqnarray}
&&\frac{\partial \Eren (\kren,\tren)}{\partial \tren} 
= \frac{1}{\pi}
\intd{\BBR}{\varepsilon^{1/2} N}
 \intd{\BBR}{\varepsilon^{1/2} N} 
\frac{\sin (\pi^3 \kren \krena \krenb \tren)}{\krena \krenb} 
\deldisc{\varepsilon^{1/2}N}(\kren-\krena-\krenb) \label{eq:modkin}\\
&& \qquad  \qquad \times 
\left[\kren \Eren (\krena,\tren) \Eren (\krenb,\tren)
- \krena \Eren (\kren,\tren) \Eren (\krenb,\tren)
- \krenb \Eren (\kren,\tren) \Eren (\krena,\tren)\right]
\, \dif \krena \, \dif \krenb \nonumber\\
\end{eqnarray}
We comment now on these predictions.
\subsection{Scaling Predictions}
The self-consistency calculation based on the weak turbulence
expansion predicts that the band of wavenumbers participating in the
earliest (three-mode) phase of energy transfer has thickness $ \Delta
k \sim \gwid = \varepsilon^{1/2} $ and associated dynamical time scale $ t
\sim \gtime = \varepsilon^{-3/2} $, provided the initial data is distributed
over a band of low wavenumbers which is comparably ($ \varepsilon^{1/2}$) 
thick.  In~\cite{jab:setfp}, 
these scaling predictions are confirmed through direct numerical
simulations.  Analytical
explanations for the $ \varepsilon^{1/2} $ width for the band of
wavenumbers which equilibrate first have been previously offered on the
basis of resonance overlap calculations~\cite{dls:lecfp} and related
considerations~\cite{bc:stfpu,lg:stsco}.  One can also construct an
argument for the $ \varepsilon^{-3/2} $ dynamical time scale from
cruder arguments.  Our intent, however, was to provide a more
systematic argument, since resonance overlap arguments can sometimes
give misleading results~\cite{rl:etnlh}.  As we discuss below, however, we
cannot yet claim to have fulfilled this goal.
\subsection{Properties of Modified Kinetic Equation}
The  modified
kinetic equation (\ref{eq:modkin}) has the following good
properties:
\begin{itemize}
\item The modified three-wave kinetic equation (\ref{eq:modkin}) 
conserves the 
harmonic energy over the three-wave time scale:
\begin{displaymath}
\frac{\dif}{\dif \tren} \intd{\Circ}{N} \Espec (\kcont,t) \, \dif \kcont
 = 0,
\end{displaymath}
More precisely, direct calculations on the modified kinetic equation
(\ref{eq:modkin}) show that the three-wave interactions conserve the harmonic
energy locally (triad by triad) for the band of wavenumbers $ \kcont
\lesssim \varepsilon^{1/2} $, while the energy in wavenumbers $ \kcont
\gg \varepsilon^{1/2} $  is predicted not to change significantly on
the time scale $ t \lesssim \varepsilon^{-3/2} $.
\item The modified kinetic equation accounts for a finite time
resonance broadening in that the usual Dirac delta function $ \delta
(\Delta \omega) $ involving the frequency difference (\ref{eq:delom})
 is instead replaced in (\ref{eq:modkin}) by a renormalized form of 
the function $ \frac{\sin (\Delta
\omega t)}{\Delta \omega} $.  While the latter does indeed converge to
the Dirac delta function in the $ t \rightarrow \infty$ limit, its
distinction from this strict limit is important for predicting
nontrivial dynamics for large but finite $ t $.  Some previous
work~\cite{cc:dqwtc,anp:kftns} offered \emph{ad hoc} modifications to the weak
turbulence theory to take into account such broadening of the
resonances, but we have endeavored to derive the quantitative form of
the finite-time resonance
broadening through systematic means.
\end{itemize}
\subsection{Deficiency in Modified Kinetic Equation}
The modified kinetic equation, however, has a serious
deficiency in that it does not
preserve nonnegativity of the energy spectrum, because the broadened
resonance kernel
$ \sin (\pi^3 \kren \krena \krenb \tren)/(\kren \krena \krenb) $ is
not sign-definite.  This indicates that, unlike standard weak
turbulence theory, the removal of the leading order secular term (along
with frequency renormalization) does
not simultaneously remove all higher order secularities active at the
first nonlinear time scale, $ t \sim \gtime = \varepsilon^{-3/2} $, in
the $ \alpha$-FPU model.
  A
similar issue arose in the problem of acoustic turbulence~\cite{vsl:sdat}.

Newell and coworkers~\cite{ljb:sfcbc,cc:dawt,acn:wtaai} have recently pointed out that the assumptions underlying the 
WT theory will often fail at either large or small scales, because the
nonlinear time scale $ \tnl $ associated with such scales may fail to be long
compared with the corresponding linear time scale $ \tlin $.  Because the
development of partial equipartition is concentrated on small
wavenumbers $ k \lesssim \gwid \sim \varepsilon^{1/2} $, one might
think that the deficiencies noted for the kinetic equation derived
above arise from this kind of breakdown in the asymptotic expansion.
We now show, however, that the simplest such explanation is insufficient.  

The linear time scale $ \tlin $ associated with a wavenumber $ \kcont
$ is just the inverse frequency $ (\omega (\kcont))^{-1} $, while the
associated nonlinear time scale $ \tnl \sim \frac{\partial \Espec
(\kcont,t)/\partial t}{\Espec (\kcont,t)} $.  From studying either the
primitive second order WT expansion (\ref{eq:newneq}) 
or the renormalized kinetic
equation, we see that for the band of active wavenumbers $ \kcont
\lesssim \varepsilon^{1/2} $, the nonlinear time scale depends on $
\kcont $ and $ \varepsilon $ as $\tnl \sim (\varepsilon \omega
(\kcont))^{-1} $.
% see 09/08/02 notes
Therefore, for $ \kcont \lesssim \varepsilon^{1/2} $,
the ratio $\tnl/\tlin \sim \varepsilon^{-1}$
is uniformly large, so the defects in the kinetic equation cannot be attributed
to this criterion elaborated in~\cite{ljb:sfcbc,cc:dawt,acn:wtaai}.

Another situation in which weak turbulence theory will
fail is in finite systems where the wavenumber spacing $ 1/N $ is not small
compared with the width of the frequency resonance, which is here
proportional to $ \gwid \sim \varepsilon^{1/2} $~\cite{acn:wti}.  Therefore, WT
theory cannot be expected to apply to the $ \alpha$-FPU model when $ N
\varepsilon^{1/2} \lesssim 1 $, but this can be avoided, at least in
principle, by choosing the size of the system $ N$ to be large enough.

At this time, we are endeavoring to specify concretely the
 source of the deficiency of the
modified kinetic equation.
\section{Derivation of Modified Kinetic Equation}
\label{sec:derkin}
Our starting point will be the following equation, which results from
a multiple-scales expansion for the moments of the mode amplitudes $
\amode $~\cite{acn:wti}:
% See 6/12/02 notes
\begin{eqnarray}
\frac{\partial \nspec (\kcont,t)}{\partial t} 
&=& -4 \varepsilon^2 \Re \sum_{\saa,\sbb=\pm}
\intd{\Circ}{N}  \intd{\Circ}{N}  
\Intexpt (\omega (\kcont) - \saa \omega (\kconta ) - \sbb
\omega (\kcontb)) \nonumber\\
& &\qquad \times 
\Ifin^{+\saa\sbb}_{\kcont\kconta\kcontb} 
\left[\Ifin^{+,\saa,\sbb}_{\kcont,\kconta,\kcontb}
\nspec (\kconta,t) \nspec (\kcontb,t) 
+ \Ifin^{-\saa,\sbb,-}_{-\kconta,\kcontb,-\kcont} \nspec (\kcontb,t) \nspec
(\kcont,t) \right. \nonumber\\
& &\qquad \qquad \qquad 
\left. + \Ifin^{-\sbb,-,\saa}_{-\kcontb,-\kcont,\kconta} \nspec(\kcont,t) \nspec
(\kconta,t)
\right] \nonumber \\
& & \qquad \times \deldisc{N} (\kcont - \kconta - \kcontb)
\expe^{i(-\omega (\kcont) + \saa \omega (\kconta)
+ \sbb \omega (\kcontb))t}  \, \dif \kconta \, \dif \kcontb\nonumber\\
& &\qquad 
+ \ldots
\label{eq:newneq}
\end{eqnarray}
Here we have used the symbol $ \Re $ to denote the real part of the
ensuing expression, and defined the function
\begin{displaymath}
\Intexpt (\varpi) = \frac{\expe^{i \varpi t} - 1}{i \varpi}.
\end{displaymath}
The ellipses denote terms which are
omitted because they are formally of higher order in $ \varepsilon $.
We will revisit the effects of these neglected terms in
future work, because  terms which are insignificant in standard
weak turbulence theory must be carefully examined to see if they
remain insignificant in our modified derivation.  

We convert Eq.~(\ref{eq:newneq}) 
into an equation for the energy spectrum through the 
relation $ \Espec (\kcont) = 2 
\omega (\kcont) \nspec (\kcont) $ and the explicit
formula (\ref{eq:ifin}) for the coupling coefficient.  Then upon
substituting the rescaling (\ref{eq:ektren}) into (\ref{eq:newneq}), we 
obtain
\begin{eqnarray}
\frac{\partial \Escale (\kren,\tren)}{\partial \tren} 
&=& - 4\varepsilon^2 \gtime
 \Re \sum_{\saa,\sbb=\pm}
\intd{\gwid^{-1}\Circ}{\gwid N}  \intd{\gwid^{-1}\Circ}{\gwid N}
  \gtime^{-1} \Intexptren (\gtime^{-1} \Delta \omega) \nonumber \\
& &\qquad \times 
\Ecoup^{+\saa\sbb}_{\gwid \kren,\gwid\krena,\gwid \krenb} 
\left[\Ecoup^{+,\saa,\sbb}_{\gwid \kren,\gwid \krena,\gwid \krenb}
\Escale (\krena,\tren) \Escale (\krenb,\tren) \right. \nonumber \\
& & \qquad \qquad \qquad
+ \Ecoup^{-\saa,\sbb,-}_{-\gwid\krena,\gwid\krenb,-\gwid\kren} 
\Escale (\krenb,\tren) \Escale (\kren,\tren)  \nonumber \\
& &\qquad \qquad \qquad 
\left. + \Ecoup^{-\sbb,-,\saa}_{-\krenb,-\kren,\krena}
\Escale (\kren,\tren) \Escale (\krena,\tren)
\right] \nonumber \\
& & \qquad \times 
\deldisc{\gwid N}(\kren-\krena+\krenb) \expe^{-i \gtime^{-1} \Delta \omega \tren} \, \dif \krena \, \dif \krenb
 \nonumber \\
& &\qquad 
+ \ldots
\label{eq:escale}
\end{eqnarray}
where
\begin{equation}
\Delta \omega \equiv \omega (\gwid \kren) -\saa \omega (\gwid \krena)
- \sbb \omega (\gwid \krenb) 
\label{eq:delom}
\end{equation}
and
\begin{displaymath}
\Ecoup_{\kcont \kconta \kcontb}^{s \saa \sbb} = \sqrt{\frac{\omega (\kcont)}{\omega (\kconta) \omega (\kcontb)}}  \Ifin_{\kcont \kconta \kcontb}^{s \saa \sbb}
= \frac{i \saa \sbb \sgn (\kdisc \kdisca \kdiscb) \omega (\kcont)}{2 \sqrt{2}}.
\end{displaymath}
We now seek particular relations between $ \gtime $, $ \gwid $, and $
\varepsilon $ so that the right hand side of (\ref{eq:escale}) is 
consistent with the assumed order unity magnitude and scale of
variation of the left hand side in the limit $ \varepsilon \rightarrow 0
$.  To this end, we
recall that $ \gwid (\varepsilon) \downarrow 0 $, $ \gtime
(\varepsilon) \rightarrow \infty $, and observe that
\begin{eqnarray*}
\omega (\kcont) &=& \omega (\gwid \kren) \sim 2 \pi \gwid |\kren| + O
(\gwid^3), \\
\Delta \omega
&=& 2 \pi \gwid (|\kren | - \saa |\krena| - \sbb |\krenb|)
- \frac{\pi^3 \gwid^3}{3} (|\kren|^3 - \saa |\krena|^3 - \sbb
|\krenb|^3) + O (\gwid^5) \\
&=& \pi \gwid (|\kren | - \saa |\krena| - \sbb |\krenb|) \\
& & \qquad \quad \times
\left[2 - 
\frac{\pi^2 \gwid^2}{3}
\left(|\kren|^2 + |\kren|(\saa |\krena| + \sbb |\krenb|)
+ \krena^2 + \krenb^2  - \saa \sbb |\krena| |\krenb|\right)\right] \\
& & \qquad 
- \pi^3 \gwid^3 
\saa \sbb |\kren| |\krena| |\krenb| + O (\gwid^5).
\end{eqnarray*}
% From 08/28/02 notes
Noting that $ \kren = \krena + \krenb $ from the Kronecker delta
function in (\ref{eq:escale}), 
the appropriate asymptotics for $ \Delta \omega $
can be decided based on the $ \saa $, $ \sbb $, and the 
signs of $ \kren $, $ \krena $, and $\krenb $.  If the signs align so
that $ |\krena + \krenb| \equiv \saa |\krena| + \sbb |\krenb| $, then
$ \Delta \omega \sim \ord (\gwid^3) $; otherwise $ \Delta \omega \sim
\ord (\gwid) $.  Moreover,
\begin{eqnarray*}
&&\expe^{-i \gtime
\Delta \omega \tren}
 \Intexptren (\gtime \Delta \omega ) \\
& & = \begin{cases} 
\gtime^{-1} \gwid^{-3} 
\left[\frac{\expe^{-i\gtime \Delta \omega \tren}-1}{
i \pi^3 \saa \sbb |\kren| |\krena| |\krenb|} + O (\gwid^2)\right]
& \mbox{ if } |\krena + \krenb| = \saa |\krena| + \sbb |\krenb|, \\ 
\gtime^{-1} \gwid^{-1} 
\left[\frac{1 - \expe^{-i \gtime \Delta \omega \tren}}{2 \pi i 
(|\kren| -  \saa |\krena| - \sbb |\krenb|)} + O (\gwid^2)\right]
& \mbox{ otherwise }.
  \end{cases}
\end{eqnarray*}
Therefore, the dominant contribution will come from the wavenumbers in
the domain such that $ |\kren| = |\krena + \krenb| = \saa |\krena| +
\sbb |\krenb| $.  We will choose the scaling functions so that $
\Delta \omega \lesssim \gtime^{-1} $ for order unity values of $ \kren $, $
\krena $, and $ \krenb $ in this dominant region, so we can simplify
further:
\begin{eqnarray*}
\expe^{-i \gtime
\Delta \omega \tren}
 \Intexptren (\gtime \Delta \omega )
&=& 
\gtime^{-1} \gwid^{-3} 
\left[\frac{\expe^{i \pi^3\gtime \gwid^3 \saa \sbb |\kren|
|\krena| |\krenb| \tren}-1}{
i \pi^3 \saa \sbb |\kren| |\krena| |\krenb|} + 
O (\gwid^2 +\gwid^5 \gtime)\right] \\
& & \mbox{ when } |\krena + \krenb| 
= \saa |\krena| + \sbb |\krenb |.
\end{eqnarray*}

Discarding subdominant contributions based on the above
considerations, 
we achieve after some analytical manipulations:
% Checked 6/22/02 and 6/12/02 against 5/25/02 notes
\begin{eqnarray*}
&&\lim_{\varepsilon \rightarrow 0}
\frac{\partial \Escale (\kren,\tren)}{\partial \tren} 
= \lim_{\varepsilon \rightarrow 0}
\frac{\varepsilon^2 \gtime }{\pi \gwid
(\varepsilon)}
\intd{\BBR}{\gwid N}
 \intd{\BBR}{\gwid N} 
\frac{\sin (\gwid^3 \gtime \pi^3 \kren \krena \krenb \tren)}{\krena
\krenb} 
\\
& &\qquad  \qquad \times 
\left[\kren \Eren (\krena,\tren) \Eren (\krenb,\tren)
- \krena \Eren (\kren,\tren) \Eren (\krenb,\tren)
- \krenb \Eren (\kren,\tren) \Eren (\krena,\tren)\right]\\
& & \qquad \qquad \times \deldisc{\gwid N}(\kren-\krena-\krenb) \, 
\dif \krena \, \dif \krenb
\end{eqnarray*}
The equating of the limit of the integral on the right hand side by
the integral of the limiting integrand can be justified through
uniform bounds on finite intervals and the relative negligibility of
the contribution from large $ \krena $ and $ \krenb $.  The latter
fact can be checked by  direct analysis or more intuitively understood from
the observation that the wavenumbers with $ \ord (1) $ magnitude $
\kcont $ (and therefore large values of $ \kren = \kcont \gwid^{-1} $)
exhibited no activity on the three-mode interaction time scale.

The gauge functions $ \gwid (\varepsilon) $ and $ \gtime (\varepsilon)
$ were introduced so that $ \Escale (\kren,\tren) $ would have
a nontrivial dynamics with order unity magnitude in the $ \varepsilon \rightarrow 0 $ limit.  We see then that a
nontrivial, finite dynamics results precisely if
$ \gwid^3 \gtime $ and $ \varepsilon^2 \gtime \gwid^{-1} $ have
finite, nonzero limits.  We thereby arrive at the unique (up to
irrelevant multiplicative constants) choice of
gauge functions (\ref{eq:scalpred}) and modified kinetic equation
(\ref{eq:modkin}) for the renormalized energy spectrum.

\section{Acknowledgements}
The authors would like to thank David Cai and Gregor Kova\v ci\v c for
helpful discussions.  PRK is supported by an NSF grant DMS-A11271, JAB
is supported by an NSF VIGRE postdoctoral research fellowship DMS-9983646, 
and YVL
is supported by NSF CAREER grant DMS-0134955 and ONR YIP grant
N00014210528.

\bibliographystyle{plain}
\bibliography{journalorig,journalmy,publishorig,publishmy,weakturb,statphys,appmath,own,turbdif,prob,dynsis}
\end{document}